\documentclass[aps,amsmath,amssymb,reprint]{revtex4-1}
\usepackage[english]{babel}
\usepackage{amsmath,amsthm}

\usepackage{dcolumn}
\usepackage{bm}

\usepackage{graphicx}
\usepackage{subfigure}
\usepackage{graphicx,picture}

\begin{document}
\preprint{APS/123-QED}

\title{Wake Potential in Strong Coupling Plasma from AdS/CFT correspondence}%

\author{Lian Liu}
\author{Hui Liu}%
 \email{tliuhui@jnu.edu.cn}
\affiliation{Department of Physics and Siyuan Laboratory, Jinan University, Guangzhou 510632, China}%

\date{\today}

\begin{abstract}
With the dielectric function computed from AdS/CFT correspondence, we studied the wake potential induced by a fast moving charge in strong coupling plasma, and compared it with the weak coupling wake potential for different particle velocities as $v=0.55c$ and $v=0.99c$.
The most prominent difference between strong and weak wake potential is that when $v=0.99c$ the remarkable oscillation due to Cerenkov-like radiation and Mach cone in weak coupling disappears in strong coupling, which implies that the plasmon mode with phase velocity lower than the speed of light dose not exist in strong coupling plasma.

\begin{description}
\item[PACS numbers]
11.10.Wx, 12.38.Mh
\end{description}
\end{abstract}


\maketitle

% ----------------------------------------------------------------
\section{Introduction}
In relativistic heavy ion collisions, high momentum partons in the quark-gluon plasma (QGP) might travel through the fireball and exchange energy with the medium. This phenomenon is usually studied from two aspects.
On one hand, the high momentum parton will loss energy to the QGP and results in jet quenching\cite{quenching1, quenching2, quenching3, quenching4, quenching5, quenching6}. On the other hand, the energy transferred by the parton will cause the unbalanced distribution of energy in the plasma and lead to wake of charge density and potential, which reflects significant properties of the medium response to the external source. The wake induced by the moving parton is what we focus on in this paper.

The wake of charge and potential in QGP were first studied by Ruppert et.al.\cite{wake1} and Chakraborty et.al.\cite{wakeQGP} within the linear response framework in Hard Thermal Loop(HTL) approximation. Then, Jiang and Li\cite{wake2} improved the calculation with HTL resummation. Rather, the wake potential in viscous QGP\cite{wake3}, collisional QGP\cite{wake4}, anisotropic QGP\cite{wake5} and collisional anisotropic QGP\cite{wake6} were discussed respectively.
However, the above analysis on wake potential in weak coupling plasma is not complete. As the QGP in Relativistic Heavy Ion Collider(RHIC) is believed to be strongly-coupled\cite{strongcoupling1, strongcoupling2, strongcoupling3}, the wake potential in strong coupling plasma is also worthy of discussion.

Since the perturbation theory fails in the discussion of strong coupling system, AdS/CFT correspondence, which connects conformal field theory(CFT) and string theory on certain background, is developed to deal with strong-coupling problems\cite{ads1, ads2, ads3, ads4, ads5}. The original and best example is the correspondence between the $\mathcal N=4$ supersymmetric Yang-Mills(SYM) theory at large N and large 't Hooft coupling and the type IIB string theory near horizon limit in $AdS_5\times S^5$ space.
In this paper, we will work in this framework and introduce a fast 'R-charged' particle\cite{Rcharge}, moving in the strong coupling $\mathcal N=4$ SYM plasma and inducing wake potential along the direction of motion. The moving R-charged particle forms an R-charge current and will be affected by the "R-photon" self-energy in which the SYM interactions dominate.
In order to compare with weak coupling case, we employ the same particle velocities as in earlier publications, say $0.55c$ and $0.99c$, which are two representative speeds for particles less and greater than the average phase velocity of weak coupling plasmon modes.

The whole paper is organized as follows. In section II, we briefly review the formalism of the wake potential in linear response theory. In section III, strong coupling dielectric function from AdS/CFT correspondence and weak coupling dielectric function from HTL approximation are presented. Section IV is the numerical comparison between strong coupling and weak coupling wake potentials for different particle velocities. The last section is the conclusion.

\section{Linear response theory}
Since we are interested in the most fundamental properties of strong coupling plasma responding to an external disturbance, we regard the strong coupling plasma as a finite, continuous, homogenous and isotropic dielectric medium, and apply the linear response theory framework to the medium disturbed by an external charge.
The properties of this medium are characterized by the dielectric tensor $\epsilon_{ij}(\omega,k)$\cite{linear1, linear2}, which can be
projected by $P^{L}_{ij}=k_ik_j/k^2$ and $P^{T}_{ij}=\delta_{ij}-k_ik_j/k^2$ due to the Lorentz violation in a thermal bath as
\begin{equation}
\epsilon_{ij}(\omega,k)=P^{L}_{ij}\epsilon_{L}(\omega,k)+P^{T}_{ij}\epsilon_{T}(\omega,k),
\end{equation}
in which $\epsilon_L$ and $\epsilon_T$ are the longitudinal and the transverse dielectric function respectively.
The longitudinal dielectric function is related to the self-energy in the medium as\cite{df1, df2}
\begin{equation}\label{dfl}
\epsilon_{L}(\omega,k)=1-\frac{\Pi_{L}(\omega,k)}{k^2},
\end{equation}
where $\Pi_L(\omega,k)=\Pi_{00}(\omega,k)$ in Coulomb gauge.

For an isotropic and homogeneous system, the charge density induced by an external charge distribution is\cite{linear1}
\begin{equation}
\rho_\textrm{ind}(\omega,k)=\left(\frac{1}{\epsilon_{L}(\omega,k)}-1\right)\rho_\textrm{ext}(\omega,k),
\end{equation}
where $\rho_\textrm{ext}$ is the external charge density.
Then the wake potential in momentum space is given by Poisson equation as
\begin{equation}\label{wake 1}
\Phi(\omega,k)=4\pi\frac{\rho_\textrm{ext}(\omega,k)}{k^2\epsilon_{L}(\omega,k)}.
\end{equation}

Now we consider a charge $Q$ moves with a constant velocity $\mathbf{v}$ along a fixed direction, the charge density associated with the charge can be expressed as
\begin{eqnarray}
\rho_\textrm{ext}&=&2\pi Q\delta(\omega-\mathbf{v}\cdot\mathbf{k}).\label{externalp}
\end{eqnarray}
Combining (\ref{wake 1}) and (\ref{externalp}), the wake potential in configuration space due to the motion of the charge becomes\cite{potential}
\begin{eqnarray}
\Phi(\mathbf{r},\mathbf{v},t)&=&2\pi Q\int\frac{\mathrm{d}^3k}{(2\pi)^3}\int\frac{\mathrm{d}\omega}{2\pi}\nonumber\\
&&\times\frac{4\pi}{k^2\epsilon_L(\omega,k)}\mathrm{e}^{i(\mathbf{k}\cdot\mathbf{r}-\omega t)}\delta(\omega-\mathbf{k}\cdot\mathbf{v}).
\end{eqnarray}

We assume the charge moves along z direction. By using cylindrical coordinate for $\mathbf{k}=(\kappa\cos\phi,\kappa\sin\phi,k_z)$ and $\mathbf{r}=(\rho,0,z)$, the wake potential can be written as\cite{wakeQGP}
\begin{eqnarray}
\Phi(\rho,z,t)&=&\frac{Q}{\pi v}\int^{\infty}_0\mathrm{d}\kappa\ \kappa J_0(\kappa\rho)\int^{\infty}_{-\infty}\mathrm{d}\omega\frac{1}{k^2\Delta(\omega,k)}\nonumber\\
&&\times\Big[\cos\left(\omega\Big(\frac{z}{v}-t\Big)\right)\mathrm{Re}\epsilon_L\nonumber\\
&&+\sin\left(\omega\Big(\frac{z}{v}-t\Big)\right)\mathrm{Im}\epsilon_L\Big],
\end{eqnarray}
where $J_0$ is the Bessel function, $k=\sqrt{\kappa^2+\omega^2/v^2}$ and $\Delta=(\mathrm{Re}\epsilon_L)^2+(\mathrm{Im}\epsilon_L)^2$.
Here we only focus on the wake potential parallel to the direction of motion where $\mathbf{v}\parallel\mathbf{r}$ and $\rho=0$. With the mode, $\omega=\mathbf{k}\cdot\mathbf{v}$, which is supported by (\ref{externalp}), the wake potential becomes
\begin{eqnarray}\label{wake 2}
\Phi(\mathbf{r},\mathbf{v},t)&=&\frac{Q}{\pi}\int^{\infty}_0\mathrm{d}k\int^1_{-1}\mathrm{d}x\Big[\frac{\mathrm{Re}\epsilon_L}{\Delta} \cos(kx|\mathbf{r}-\mathbf{v}t|)\nonumber\\
&&+\frac{\mathrm{Im}\epsilon_L}{\Delta} \sin(kx|\mathbf{r}-\mathbf{v}t|)\Big],
\end{eqnarray}
where $x=\cos\theta$ is the cosine of the angel between $\mathbf{k}$ and $\mathbf{v}$.
Apparently the longitudinal dielectric function $\epsilon_L$ is crucial in the calculation of wake potential.

\section{Dielectric function}
\subsection{Dielectric function  AdS/CFT correspondence}
For simplicity, we will examine an "electromagnetism-like" probe in the strong coupling plasma. Following\cite{Rcharge}, we add a conserved U(1) current corresponding to the U(1) subgroup in the model consist of $\mathcal N=4$ SYM gauge bosons plus R-charge Wyle fermions. The fermions, on one hand couple to the non-Abelian SYM gauge field with "color" charge, on the other hand also couple to the fast moving external current with an "electromagnetism-like" R-charge. In this way, when the R-charge particle gets through the strong coupling plasma, the response from the medium will be owing to a R-photon self-energy inside which the SYM interaction prevails.
The Lagrangian of this R-charge model is
\begin{equation}
\mathcal{L}=\mathcal{L}_\textrm{SYM}+\mathcal{L}_\textrm{int}-
\frac{1}{4}F^2_{\mu\nu}-\bar{\ell}\gamma^\mu(\partial_\mu-ieA_\mu)\ell,
\end{equation}
where $\mathcal{L}_\textrm{SYM}$ is the lagrangian of $\mathcal N=4$ SYM theory, $\mathcal{L}_\textrm{int}$ is the interacting term of R-current in $\mathcal N=4$ SYM theory, $F_{\mu\nu}$ is the field tensor of R-photon, $\ell$ is the Wyle fermions and $A_\mu$ is the R-photon field. The last term gives the coupling vertex of fermions and R-photon with $e$ the coupling constant.

In AdS/CFT duality, the $\mathcal N=4$ SYM is dual to the type IIB string theory in $AdS_5\times S^5$ space, which is governed by the metric
\begin{eqnarray}\label{metric}
\mathrm{d}s^2_{10}&=&\frac{(\pi T R)^2}{u}(-f(u)\mathrm{d}t^2+\mathrm{d}x^2+\mathrm{d}y^2+\mathrm{d}z^2)\nonumber\\
&&+\frac{R^2}{4u^2f(u)}\mathrm{d}u^2+R^2\mathrm{d}\Omega^2_5,
\end{eqnarray}
where $T$ is the Hawking temperature, and $f(u)=1-u^2$. The horizon corresponds to $u=1$ and the spatial infinity corresponds to $u=0$.
The five-dimensional Maxwell equation in the background (\ref{metric}) is
\begin{equation}
\partial_\mu \sqrt{-g}F^{\mu\nu}=0,
\end{equation}
where $\sqrt{-g}$ is the metric norm. With gauge condition $A_u=0$, one obtains the equations of motion
\begin{eqnarray}
\hat{\omega}A'_t+\hat{k}fA'_z=0,\label{EOM1}\\
A''_t-\frac{1}{uf}(\hat{k}^2A_t+\hat{\omega}\hat{k}A_z)=0,\label{EOM2}\\
A''_z+\frac{f'}{f}A'_z+\frac{1}{uf^2}(\hat{\omega}^2A_z+\hat{\omega}\hat{k}A_t)=0,\\
A''_\alpha+\frac{f'}{f}A'_\alpha+\frac{1}{uf}(\frac{\hat{\omega}^2}{f}-\hat{k}^2)A_\alpha=0,
\end{eqnarray}
where $\hat{\omega}=\omega/2\pi T$ and $\hat{k}=k/2\pi T$ are dimensionless energy and momentum respectively, $\alpha$ stands for either $x$ or $y$, and the derivatives are with respect to $u$.

The Minkowskian Green's function is defined in the usual way
\begin{equation}
\Pi^{R}_{\mu\nu}(\omega,\mathbf{k})=-ie^2\int\mathrm{d}^4x\mathrm{e}^{-ik\cdot x}\theta(t)\langle[j_\mu(x),j_\nu(0)]\rangle,
\end{equation}
where $j_{\mu}=\bar{\ell}A_{\mu}\ell$ is the R-charge current.
From the relevant part of the action for R-photon field and the prescription for Minkowskian Green's function formulated in \cite{adsGreen}, the strong coupling longitudinal R-photon self-energy is obtained as
\begin{equation}\label{Green's function}
\Pi_{L}^s(\omega,k)=\frac{e^2N^2T^2}{8}\frac{\hat{k}^2A'_t}{ufA''_t}\bigg|_{u \to 0},
\end{equation}
in which $N$ is the number of quark colors, and the component $A_t$ satisfies the differential equation
\begin{equation}\label{At}
A'''_t+\frac{(uf)'}{uf}A''_t+\frac{\hat{\omega}^2-\hat{k}^2f}{uf^2}A'_t=0,
\end{equation}
which is obtained by combining (\ref{EOM1}) and (\ref{EOM2}).

The strong coupling self-energy(\ref{Green's function}) suffers a logarithm divergence, which corresponds to the ultraviolet divergence in zero-temperature quantum field theory. To control this divergence and compare with weak coupling case, we subtract the zero-temperature contribution and absorb it into the renormalized coupling constant, then the non-divergent temperature-dependent photon self-energy yields\cite{cxm}
\begin{equation}\label{self energy}
\Pi^s_L(\omega,k)=\frac{e^2N^2}{8}\Big[\frac{T^2\hat{k}^2A'_t}{ufA''_t}\Big|_{u\to 0}-\frac{k^2A'_{t0}}{(2\pi)^2xA''_{t0}}\Big|_{x\to 0}\Big],
\end{equation}
where $A_{t0}$ in the second term is the solution of the zero temperature differential equation
\begin{equation}\label{zeroT}
A_{t0}'''+\frac{1}{x}A_{t0}''+\frac{\omega^2-k^2}{(2\pi)^2x}A_{t0}'=0,
\end{equation}
in which $x=u/T^2$ is the zero temperature variable substitution, and the derivative of $A_{t0}$ is with respect to $x$. In zero temperature limit, the solution of (\ref{zeroT}) is
\begin{equation}
A_{t0}'=K_0(\frac{\sqrt{x(q^2-\omega^2})}{\pi}),
\end{equation}
where $K_0$ is the modified Bessel function of the second kind.
Inserting (\ref{self energy}) into (\ref{dfl}), one will obtain the strong coupling dielectric function.

\subsection{Dielectric function in weak coupling}
The weak coupling dielectric function can be calculated from the bare one-loop self-energy or its resummation. In high temperature limit, the HTL approximation is applied to deal with the one-loop gluon self-energy, which separates integral momenta into "hard" and "soft" momenta. With hard loop momenta, which is order of $T$, and soft exterior line momenta, which is order of $eT$, the gluon self-energy is analytic and gauge independent.
Rather, the longitudinal dielectric functions of this approximation reads as\cite{df1, df2}
\begin{equation}\label{htll}
\epsilon_L^\textrm{w}(\omega,k)=1+\frac{m^2_D}{k^2}\Big[1-\frac{\omega}{2k}\Big(\mathrm{ln}\Big|\frac{\omega+k}{\omega-k}\Big|-i\pi\Theta(k^2-\omega^2)\Big)\Big],
\end{equation}
where $m_D^2=e^2T^2/3$ is the Debye mass.

\section{Numerical results}
In this section, we will compute the wake potential in strong coupling plasma and compare it with that in weak coupling case. The wake potential scaled with $m_D$ is rewritten as
\begin{eqnarray}\label{wake potential}
\frac{\Phi(\mathbf{z},\mathbf{v},t)}{Q m_D}&=&\frac{1}{\pi}\int^{\infty}_0\mathrm{d}k\int^1_{-1}\mathrm{d}x\Big[\frac{\mathrm{Re}\epsilon_L}{\Delta} \cos(kx|\mathbf{r}-\mathbf{v}t|m_D)\nonumber\\
&&+\frac{\mathrm{Im}\epsilon_L}{\Delta} \sin(k x|\mathbf{r}-\mathbf{v}t|m_D)\Big]\nonumber\\
&\equiv&\Phi_1+\Phi_2,
\end{eqnarray}
where $\Phi_1$ and $\Phi_2$ is the first and second terms of the integrals respectively.

Inserting (\ref{self energy}) into (\ref{dfl}) and working out the integral in (\ref{wake potential}) numerically with the quark colors $N=3$ and the finite temperature $T=0.2$GeV, we obtain the wake potential in strong coupling plasma for particle velocity $v=0.55c$ and $v=0.99c$ which are shown in Fig.\ref{sawv}. For comparison, we present the weak coupling wake potential with HTL dielectric function (\ref{htll}) in the same panel.
We find the differences between strong coupling and weak coupling are reflected in the following three aspects:

\begin{figure}[ht]
\centering
\subfigure[]{
\label{sawv5}
\includegraphics[width=0.4\textwidth]{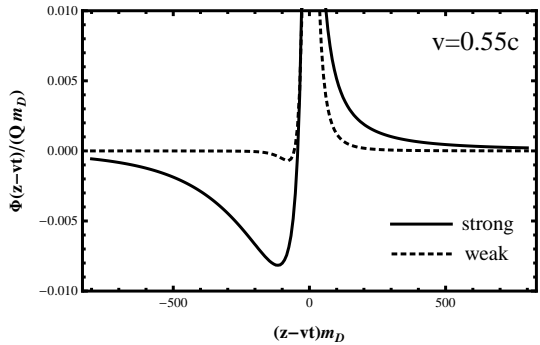}}
\subfigure[]{
\label{sawv9}
\includegraphics[width=0.4\textwidth]{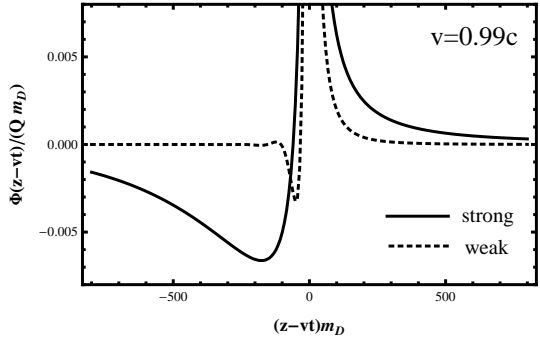}}
\subfigure[]{
\label{inter}
\includegraphics[width=0.38\textwidth]{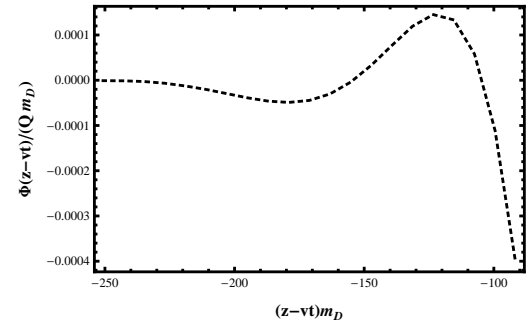}}
\caption{(a): Strong coupling (solid) and weak coupling (dashed) wake potential along the direction of the moving charge for $v=0.55c$. (b): Same as (a) but for $v=0.99c$. (c): Detail of weak coupling potential in (b).}
\label{sawv}
\end{figure}
(i)For $v=0.55c$, the strong coupling wake potential behaves like the weak coupling wake potential but with deeper negative minimum in the backward direction.

Fig.\ref{sawv5} shows the comparison of strong (solid) and weak (dashed) wake potentials for the charged particle with $v=0.55c$. In the forward direction ($(z-vt)m_D>0$), they are both like a modified Coulomb potential, due to the screening effect. Notice the strong screening curve is above the weak one, we find it is consistent with our earlier study\cite{ll} on Debye screening.
In the backward direction ($(z-vt)m_D<0$), the strong coupling and weak coupling wake potentials are both Lennard-Jones type potential, which has a short range repulsive part as well as a long range attractive part and hence form a negative minimum. However, the depth of the minimum in strong coupling is much greater than that in weak coupling, indicating that a bound state might be more easily formed along behind the fast moving charge.

(ii)For $v=0.99c$, the most prominent feature of weak coupling wake potential -- the oscillation in the backward direction -- smears out in the strong coupling case.

In Fig.\ref{sawv9}, we display the $v=0.99c$ wake potentials of strong coupling (solid) and weak coupling (dashed). The backward oscillation in the weak coupling wake potential is clearly visible from Fig.\ref{inter}. The physics explanation for this oscillation is the Cherenkov-like radiation of a fast moving charge with the speed greater than the average speed of plasmon mode. However, we carefully examined the backward direction in strong coupling wake potential, but failed to find such oscillation, i.e., the backward wake potential in strong coupling is still Lennard-Jones type potential. We check the charge velocity  more close to the speed of light, but the oscillation is still absent. The disappearance of the oscillation might be a characteristic of strong coupling plasma, indicating that the plasmon mode with phase velocity lower than the speed of light dose not exist in strong coupling limit. This character was hinted in our previous paper\cite{cxm} on the dielectric function from  AdS/CFT correspondence, in which we found the dielectric function had only a very sharp singularity on mass shell which indicated the speed of plasmon mode in strong coupling plasma is almost equal to the speed of light.

\begin{figure}[ht]
\centering
\subfigure[]{
\label{raihtl}
\includegraphics[width=0.4\textwidth]{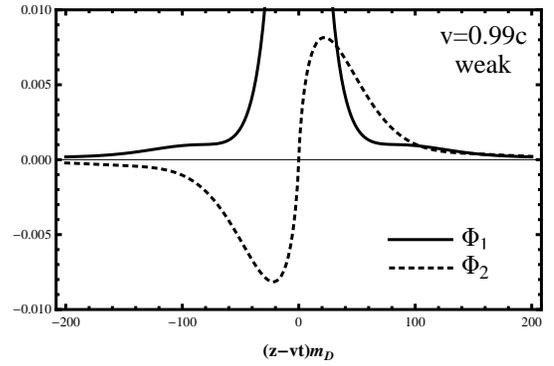}}
\subfigure[]{
\label{raiads}
\includegraphics[width=0.4\textwidth]{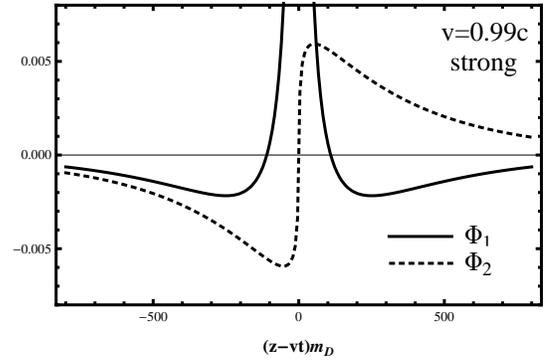}}
\caption{(a): $\Phi_1$ (solid) and $\Phi_2$ (dashed) in weak coupling wake potential along the direction of the moving charge for $v=0.99c$. (b): Same as (a) but for strong coupling wake potential. }
\label{rai}
\end{figure}

In order to get a further understanding of the oscillation disappearance in wake potential, we present the first and the second terms of the potentials i.e., $\Phi_1$ and $\Phi_2$ in (\ref{wake potential}), along the direction of the moving charge for $v=0.99c$ in Fig.\ref{rai}.
Fig.\ref{raihtl} describes $\Phi_1$ and $\Phi_2$ of weak coupling potential. $\Phi_1$ is symmetric under inversion of $(z-vt)\to -(z-vt)$, asymptotically approaching zero from above. While $\Phi_2$ is antisymmetric under inversion of $(z-vt)\to -(z-vt)$, exhibiting a negative minimum in the backward direction and then asymptotically approaching zero from below. Therefore, the competition between $\Phi_1$ and $\Phi_2$ results in the oscillation in weak coupling potential.
Fig.\ref{raiads} is the strong coupling case. Notice that $\Phi_1$ and  $\Phi_2$ in the strong coupling are both approaching zero from below in the backward direction, so that they do not compete with each other but produce a deeper negative minimum in this direction.

\begin{figure}[ht]
\centering
\subfigure[]{
\label{adsv}
\includegraphics[width=0.4\textwidth]{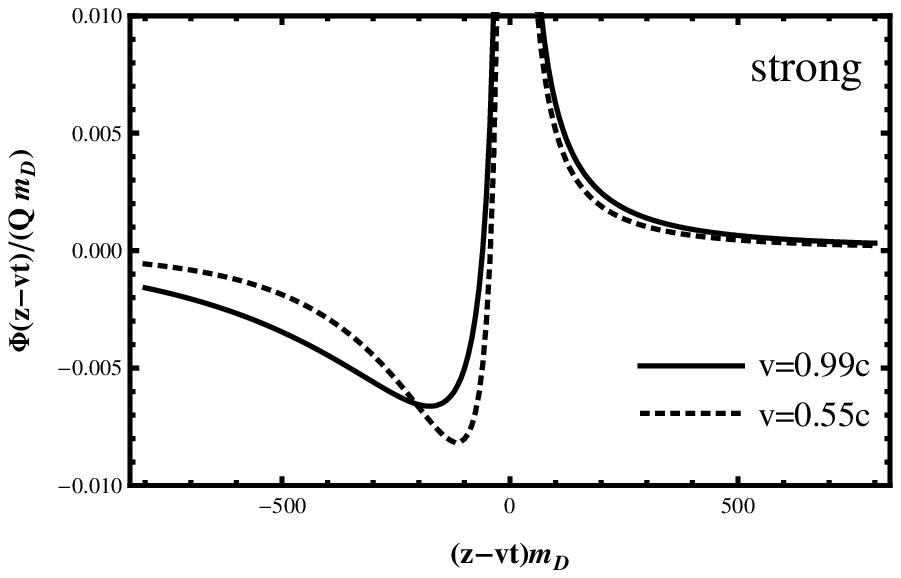}}
\subfigure[]{
\label{htlv}
\includegraphics[width=0.4\textwidth]{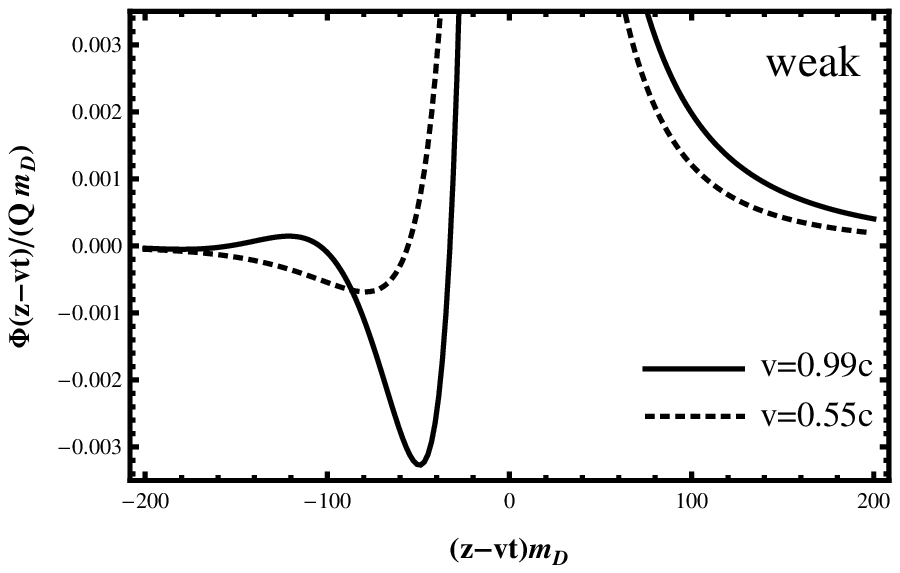}}
\caption{(a): Strong coupling wake potential along the direction of the moving charge for $v=0.55c$ (dashed) and $v=0.99c$ (solid). (b): Same as (a) but for weak coupling.}
\label{comparev}
\end{figure}
(iii)The depths and positions of negative minimum for strong and weak coupling wake potential in the backward direction display opposite variation tendencies with the increase of particle velocity, and the wake potential in strong coupling is less sensitive to the particle velocity than the weak coupling wake potential.

To make a comprehensive comparison, we regroup the potentials in strong coupling for $v=0.55c$ in Fig.\ref{sawv5} and $v=0.99c$ in Fig.\ref{sawv9} together into Fig.\ref{adsv}, and the potentials in weak coupling for $v=0.55c$ and $v=0.99c$ together into Fig.\ref{htlv}.

In the forward direction, the wake potentials for $v=0.55c$ in Fig.\ref{adsv} and Fig.\ref{htlv} both drop faster than those for $v=0.99c$, which indicates that the increase of velocity leads to the decrease of screening.
In the backward direction of Fig.\ref{adsv}, with the increase of velocity, the depth of negative minimum decreases and the position of the minimum shifts away from the zero. While these variation tendencies are both inverse to the weak coupling wake potential in Fig.\ref{htlv}. Furthermore, we find the wake potential in strong coupling is less sensitive to the velocity than that in weak coupling, as the depth of negative minimum of wake potential in strong coupling decreases about 20\% while that in weak coupling increases about 400\% when the velocity changes from $v=0.55c$ to $v=0.99c$.

\section{Conclusion}
With the dielectric function obtained from AdS/CFT correspondence, we investigate the strong coupling wake potentials induced by the fast moving charged particles with velocities $v=0.55c$ and $v=0.99c$ respectively, and compare them with those in weak coupling case.
We find that for $v=0.99c$, the remarkable oscillation akin to cherenkov-like radiation and Mach cone in weak coupling wake potential does not appear in the strong coupling wake potential, which may indicate that the phase velocity of strong coupling plasmon mode will not be lower than the speed of light. Besides this prominent difference, the wake potentials in strong and weak coupling are qualitatively similar except for some detailed discrepancies. For example, when $v=0.55c$ the strong coupling wake potential shows a deeper negative minimum in the backward direction than that in the weak coupling wake potential, the depths and positions of the negative minimum for strong and weak coupling wake potentials display opposite variation tendencies with the increase of particle velocity, and the wake potential in strong coupling is less sensitive to the particle velocity than the weak coupling wake potential.

\begin{acknowledgments}
We thank B.-F Jiang and J.-R Li for their helpful discussions and suggestions. This work is supported by National Natural Science Foundation of China under Grant No. 11405074.
\end{acknowledgments}

%-----------------------------------------------------------------

\bibliography{wakeliu}
\end{document}